\documentclass{article}
\usepackage{amsmath,graphicx}
\usepackage{booktabs}

\usepackage{amsmath,epsfig,amssymb,amsthm,url,amsfonts}
\usepackage{algorithm}
\usepackage{algpseudocode}

\usepackage{multirow}
\usepackage{mdframed}
\usepackage{url}
\usepackage{enumitem}
\usepackage[makeroom]{cancel}
\usepackage{dblfloatfix}
\usepackage{array}
\usepackage{epstopdf}
\usepackage{float}
\usepackage{subfig}
\usepackage{breqn}
\usepackage{xcolor}
\usepackage{tabularx}
\usepackage[printonlyused,withpage]{acronym}
\usepackage{balance}
\usepackage{math}
\usepackage[backend=bibtex,style=ieee,minbibnames=1,maxbibnames=4]{biblatex}
\usepackage{spconf}
\usepackage[keeplastbox]{flushend}

\usepackage{tikz}
\usetikzlibrary{bayesnet,calc,tikzmark}
\usetikzlibrary{shapes.geometric}
\tikzset{  net/.style={draw,trapezium,trapezium angle=75,shape border rotate=270} }
\tikzset{  rnn/.style={draw,rectangle} }

\theoremstyle{break}

\theoremstyle{definition}

\title{The impact of removing head movements on audio-visual\\speech enhancement}
\name{%
Zhiqi Kang$^1$, %
Mostafa Sadeghi$^2$, %
Radu Horaud$^1$, \thanks{This work has been partially supported by the H2020 SPRING project \#871245 and by the Multidisciplinary Institute of Artificial Intelligence (MIAI) ANR-19-P3IA-0003.}%
Xavier Alameda-Pineda$^1$, %
Jacob Donley$^3$ and %
Anurag Kumar$^3$%
}
\address{%
$^1$Inria Grenoble Rh\^{o}ne-Alpes \& Univ. Grenoble Alpes, France, 
$^2$Inria Nancy Grand-Est, France \\ %
$^3$Reality Labs Research, Redmond WA, USA\vspace{-3mm}}

\acrodef{STOI}{short-time objective intelligibility}
\acrodef{SE}{speech enhancement}
\acrodef{STFT}{short-time Fourier transform}
\acrodef{PSD}{power spectral density}
\acrodef{NMF}{nonnegative matrix factorization}
\acrodef{AV}{audio-visual}
\acrodef{DNN}{deep neural network}
\acrodefplural{DNNs}{deep neural networks}
\acrodef{VAE}{variational auto-encoder}
\acrodefplural{VAEs}{variational auto-encoders}
\acrodef{CVAE}{conditional variational auto-encoder}
\acrodefplural{CVAEs}{conditional variational auto-encoders}
\acrodef{A-VAE}{audio VAE}
\acrodef{V-VAE}{visual VAE}
\acrodef{AV-CVAE}{audio-visual CVAE}
\acrodef{ROI}{region of interest}
\acrodef{MCMC}{Markov Chain Monte Carlo}
\acrodef{EM}{expectation-maximization}
\acrodef{VEM}{variational expectation-maximization}
\acrodef{MCEM}{Monte Carlo expectation-maximization}
\acrodef{TF}{time frequency}
\acrodef{ELBO}{evidence lower bound}
\acrodef{ROI}{region of interest}
\acrodef{LR}{Living Room}
\acrodef{SDR}{signal-to-distortion ratio}
\acrodef{PESQ}{perceptual evaluation of speech quality}
\acrodef{ASE}{audio speech enhancement}
\acrodef{VSE}{visual speech enhancement}
\acrodef{AVSE}{audio-visual speech enhancement}
\acrodef{SNR}{signal-to-noise ratio}
\acrodefplural{SNRs}{signal-to-noise ratios}
\acrodef{LSTM}{long short-term memory}
\acrodef{HMM}{hidden Markov model}
\acrodef{SwVAE}{switching variational auto-encoder}
\acrodef{GAN}{generative adversarial network}

\begin{document}
%
\maketitle
\begin{abstract}
This paper investigates the impact of head movements on audio-visual speech enhancement (AVSE). Although being a common conversational feature, head movements have been ignored by past and recent studies: they challenge today's learning-based methods as they often degrade the performance of models that are trained on clean, frontal, and steady face images. To alleviate this problem, we propose to use robust face frontalization (RFF) in combination with an AVSE method based on a  variational auto-encoder (VAE) model. We briefly describe the basic ingredients of the proposed pipeline and we perform experiments with a recently released audio-visual dataset. In the light of these experiments, and based on three standard metrics, namely STOI, PESQ and SI-SDR, we conclude that RFF improves the performance of AVSE by a considerable margin.\footnote{https://team.inria.fr/robotlearn/head-movements-avse/}
\end{abstract}
\section{Introduction}
\label{sec:introduction}
It has long been established that speech communication is multimodal. In particular, vision provides an alternative representation of some of the information that is present in the audio, with the advantage that it is not affected by acoustic noise. This has led to speech recognition and speech enhancement systems that combine audio and visual features to achieve robust performance in noise \cite{michelsanti2021overview}.
While there is a long history of research in audio-visual speech enhancement (AVSE), the most performant methods are based on \acp{DNN}. Most \ac{DNN}-based methods for AVSE are \textit{supervised} \cite{hou2018audio,AfouCZ18,michelsanti2021overview,GabbSP18}: they learn a direct mapping from the noisy speech and clean visual inputs onto the clean speech output. This mode of doing requires large-scale datasets with a high variety of noise types and \ac{SNR} levels, which suffers from poor generalization to unseen noise environments. Alternatively, \textit{unsupervised} AVSE methods do not necessitate noise signals for training, e.g. \cite{sadeghi2020audio,sadeghi2021mixture}. They use \acp{VAE}~\cite{KingW14} to model the generative process of audio-visual speech using multimodal datasets of clean audio and  visual speech. Then, this model is combined with an audio noise model, e.g.\ \ac{NMF}, to perform speech enhancement from noisy audio and visual speech. Thanks to the independence of the noise type, the unsupervised methods exhibit better generalization than their supervised counterpart.

Nevertheless, to date, both supervised and unsupervised AVSE methods assume clean visual information, namely they use lip regions that are cropped from clean, frontal and steady face images. In practice, the visual data are corrupted by various sources of perturbation, such as partial occlusion with an object and by head movements. The performance of existing AVSE systems rapidly degrades in the presence of corrupted visual information. Recently, there have been a few attempts to incorporate knowledge about the quality of the visual data at hand and to ignore the visual input whenever it cannot be properly exploited \cite{afouras2019my,SadeA19a,sadeghi2021switching}. Here, we investigate AVSE models that put audio and visual information on an equal footing and that can deal with both noisy audio signals and noisy lip movements -- a largely uninvestigated topic. We propose to remove rigid head movements using a face frontalization method that preserves non-rigid facial deformations.

The rest of the paper is organized as follows. Section~\ref{sec:avvae} reviews the VAE-based AVSE model \cite{sadeghi2020audio}. Section~\ref{sec:frontalization} sumarizes the face frontalization method of \cite{kang2021robust} . Section~\ref{sec:exp} describes the experiments and discusses the results.
  
\section{Audio-Visual Variational Autoencoder}
\label{sec:avvae}
In this section, we briefly summarize the conditional VAE (CVAE) based AVSE  model of \cite{sadeghi2020audio}, denoted by AV-CVAE. The whole framework consists of two steps: training and testing (inference). In the first step, a prior distribution for clean speech is learned from the concatenation of a clean audio signal with an embedding of associated lip images. The second step infers clean speech from the noisy-speech and lip-embedding inputs: the learned prior distribution is combined with a noise model, whose parameters together with the clean speech ones are estimated following a variational expectation-maximization (VEM) procedure.

\begin{figure}[t]
    \centering
    \includegraphics[width=\linewidth]{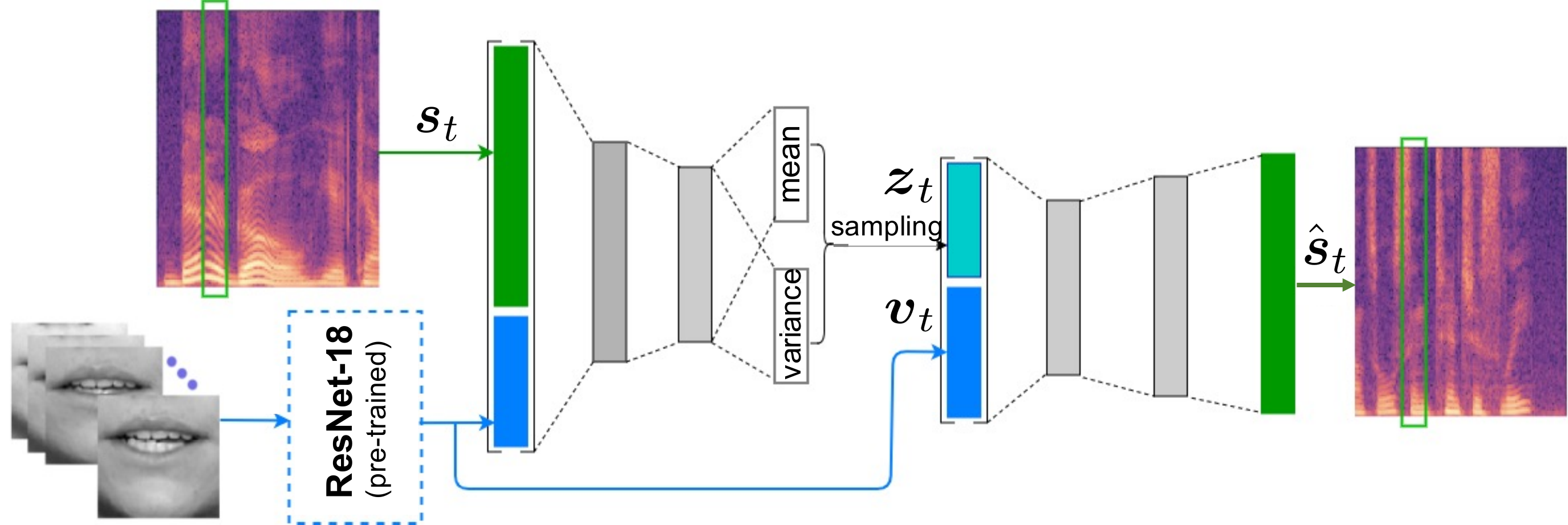}
    
    \caption{AV-CVAE model with a pre-trained ResNet-18 network as visual feature extractor.}\label{fig:av-cvae}
  
  \end{figure}

Given a dataset of complex-valued short-time Fourier transform (STFT) frames of a clean-speech signal, denoted $\svect_{t}\in\mathbb{C}^F$, and the corresponding lip embedding obtained from a lip bounding-box cropped from the image of a speaker face, denoted $\vvect_t \in \mathbb{R}^M$, a latent-variable generative model is trained using the VAE framework. This involves defining a parametric distribution for the likelihood $p_{\Theta}(\svect_{t}|\zvect_t, \vvect_t)$, and a parametric prior distribution for the latent code $\zvect_{t}\in\mathbb{R}^L$, $L\ll F$, $p_{\Gamma}(\zvect_t|\vvect_t)$. These distributions are implemented by some deep neural network architectures, whose parameters, \{$\Theta, \Gamma$\}, are learned following an amortized variational inference \cite{KingW14}, where an encoder network is introduced to approximate the intractable posterior distribution of the latent codes. Fig.~\ref{fig:av-cvae} illustrates the AV-CVAE architecture. The main difference between this architecture and the one proposed in \cite{sadeghi2020audio,sadeghi2021mixture} is the presence of a ResNet backbone from a pretrained model specialized for lip reading~\cite{martinez2020lipreading}.

With the parametric prior distribution for clean speech being learned, one considers an observation model as $\ovect_{t} = \svect_{t}+\bvect_{t}$, in which $\ovect_{t}\in\mathbb{C}^F$ and  $\bvect_{t}\in\mathbb{C}^F$ denote observed speech and noise, respectively. Considering an NMF-based model for noise, and combining with the speech model, the set of NMF parameters are then learned by a variational inference procedure. Once learned, the clean speech estimate $\hat{\svect}_{t}$ is obtained via a probabilistic Wiener filtering. More details can be found in \cite{sadeghi2020audio}.

\section{Robust Face Frontalization}
\label{sec:frontalization}
\begin{figure}[t]
\includegraphics[width=0.16\columnwidth]{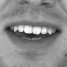}
\includegraphics[width=0.16\columnwidth]{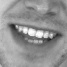}
\includegraphics[width=0.16\columnwidth]{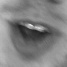}
\includegraphics[width=0.16\columnwidth]{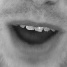}
\includegraphics[width=0.16\columnwidth]{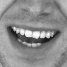}
\includegraphics[width=0.16\columnwidth]{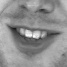}
\\[\smallskipamount]
\includegraphics[width=0.16\columnwidth]{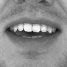}
\includegraphics[width=0.16\columnwidth]{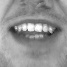}
\includegraphics[width=0.16\columnwidth]{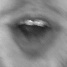}
\includegraphics[width=0.16\columnwidth]{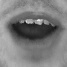}
\includegraphics[width=0.16\columnwidth]{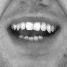}
\includegraphics[width=0.16\columnwidth]{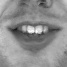}
\\ 
\vspace{-10pt}
\caption{An example of  the effect of face frontalization on the lip regions cropped from the images of a head-moving speaker. Top: lip regions cropped from the original images (with head motions). Bottom: lip regions cropped from the frontalized images (head motions removed).}\label{fig:lips}
\end{figure}

We briefly describe a method for removing head movements with the goal to provide frontal and steady lip regions to the AV-CVAE model outlined above, e.g. Figure~\ref{fig:lips}.
The core idea of the robust face frontalization (RFF) method that we recently proposed \cite{kang2021robust}, is to estimate the pose (scale $\sigma$, 3D rotation $\Rmat$ and 3D translation $\tvect$) and the 3D deformable shape $\svect$ of an arbitrarily-viewed input face, and to warp it onto a frontally viewed synthesized face. The main feature of the method is to perform pose and shape-deformation estimation sequentially. This way of doing enables a better estimation of the pose parameters in the presence of facial expressions, e.g. lip movements.

The pose is estimated by aligning two 3D points sets: a set of \textit{observed} facial landmarks extracted from the input face, $ \{ \Xvect_j \}_{j=1}^J \subset \mathbb{R}^{3}$, and a set of \textit{model} landmarks  associated with a neutral and frontal view of a mean face, $ \{ \Zvect_j \}_{j=1}^J \subset \mathbb{R}^{3}$. Because the landmark locations of the input face are inherently affected by detection errors as well as by \textit{non-rigid facial deformations}, it is suitable to use a robust rigid-parameter estimation technique. For this purpose, we assume that the errors between the model and frontalized landmarks, i.e. $\evect_j = \Zvect_j - \sigma\Rmat\Xvect_j - \tvect$, are samples of a random variable drawn from the Student t-distribution -- a heavy tailed pdf that is able to deal with both Gaussian (inliers) and non-Gaussian (outliers) noise in the data, by assigning a weight random variable to each observed landmark \cite{forbes2014new}. The corresponding EM algorithm alternates between the estimation of (i)~the weight posteriors, (ii)~the pdf parameters and (iii) the rigid parameters. At convergence, EM assigns high weight posteriors to landmark pairs that are linked by a rigid transformation and low weights to landmark pairs that are affected by detection errors or by non-rigid facial deformations. 

Next, the exact shape of the face is estimated by fitting a 3D morphable model (3DMM) to the frontalized 3D landmarks denoted $\{\Yvect_j\}_{j=1}^J$, with $\Yvect_j= \sigma^\ast\Rmat^\ast\Xvect_j+\tvect^\ast$, where $^\ast$ indicates the optimal parameter values previously computed. We use a linear deformation model which consists of a 3D mesh whose vertices are parameterized by a low-dimensional embedding $\svect$. Once the face embedding is estimated, a frontal dense depth map of the face is built, which in turn enables the input face to be warped onto the frontalized one. The final step is to crop a lip region from the frontalized face.

The face alignment method of \cite{bulat2017far} (being one of the best performing 3D face alignment methods to date), is used to extract 3D landmarks. The Student EM algorithm is described in detail in \cite{kang2021robust}. All the computations inside EM are in closed form, with the exception of the rotation. The latter is parameterized by a unit quaternion which allows a compact representation of the rotation and the use of a sequential least-squares programming (SLSQP) solver in combination with a root-finding software package \cite{kraft1988software}. Without loss of generality, we use a linear deformation model, namely the publicly available Basel Shape Model \cite{paysan20093d}. This provides registered face scans of 200 identities: 200 frontal scans with a neutral expression as well as 200 expressive scans. Each scan is described with $N=53490$ vertices. Among these vertices, $J=68$ are used for pose and shape estimation,  $\{ \Zvect_j \}_{j=1}^J$. The low-dimensional embedding is set to $K=200$.

\section{Experiments}
\label{sec:exp}

\begin{table*}[h]
\centering
	\caption{Average STOI, PESQ, SI-SDR values.}
\resizebox{\textwidth}{!}{
\begin{tabular}{|l|c|c|c|c|c||c|c|c|c|c||c|c|c|c|c|}
\hline
  Measure & \multicolumn{5}{c||}{STOI $[0,1]\uparrow$} & \multicolumn{5}{c||}{PESQ $[-0.5,4.5]\uparrow$} & \multicolumn{5}{c|}{SI-SDR (dB) $\uparrow$} \\
\hline
{SNR (dB)} & {-10} & {-5} & {0} & {5} & {10} & {-10} & {-5} & {0} & {5} & {10} & {-10} & {-5} & {0} & {5} & {10}  \\ \hline\hline
Noisy audio input & 0.40 & 0.53 & 0.66 & 0.78 & \textbf{0.86} & 0.90        & 1.24 & 1.67      & 2.05       & 2.42   & -15.92 & -10.62 & -5.44 & -0.40 & 4.60                   \\ \hline
A-VAE \cite{Leglaive_MLSP18} & 0.41 & 0.56& 0.70 & \textbf{0.79} & 0.85 & 0.93         & 1.51        & 2.02       & 2.43       & 2.73 & -7.01 & -0.29 & 5.08 & \textbf{9.41} & \textbf{12.74}  \\ \hline
AV-CVAE \cite{sadeghi2020audio}  & 0.42 & 0.57 & 0.69 & \textbf{0.79} & 0.84  & 1.02 & 1.56 & 2.06 & 2.42 & 2.73  &      -6.96  & -0.04                  & 5.01                     &  9.06                    & 12.25 \\ \hline
Res-AV-CVAE-WithHM & 0.41 & 0.55 & 0.67 & 0.77 & 0.83 &   1.02     &   1.53                 &  1.99                    &  2.35                    &  2.70 & -7.84 &  -0.60 & 4.68 & 8.81 & 12.30 \\ \hline
Res-AV-CVAE-DA-ST-GAN \cite{zhou2020rotate}  & 0.40 &  0.55 & 0.68 & 0.78 & 0.84 & 1.01  & 1.54  & 2.01 & 2.39 & 2.72 &     -7.92   &   -1.14                & 4.13                    &        9.27              & 11.77\\ \hline
Res-AV-CVAE-DA-GAN \cite{yin2020dual}  & 0.39  & 0.55 & 0.66 & 0.68 & 0.72 & 0.76  &  1.42 & 1.87 & 1.66 & 1.96 & -9.08        &  -0.45                  & 3.88                     &  4.55            & 5.23 \\ \hline
Res-AV-CVAE-RFF \cite{kang2021robust} & \textbf{0.43} & \textbf{0.58}& \textbf{0.71}& \textbf{0.79} & 0.85 & \textbf{1.12}                     & \textbf{1.69}                   & \textbf{2.13}                     & \textbf{2.48}                     & \textbf{2.77}  & \textbf{-6.30} & \textbf{0.10} & \textbf{5.24}& 9.30 & 12.60 \\ \hline
\end{tabular}}
\label{tab:measures}\vspace{-3mm}
\end{table*}

All the experiments reported below use the
MEAD dataset~\cite{wang2020mead} which contains short videos of talking faces with large-scale facial expressions. For all 46 publicly available participants, there are recordings of eight different emotions at three different intensity levels and seven camera viewpoints. Many participants have natural head motions, which challenges state-of-the-art AVSE. Among all videos, we select the videos of all emotion categories taken at the frontal view and at the level 3 (the highest) of emotion intensity. These high-intensity emotions are associated with large head movements and exaggerated lip motions, thus allowing to assess the effect of head movements. In total, there are around 5 hours of videos for training, 0.7 hours for validation and 0.7 hours for testing.

We process the input videos with three different frontalization methods to compare their effectiveness of removing head movements and hence of improving the quality of the speech output. 
We consider the following three methods. The first one \cite{kang2021robust}, denoted RFF and outlined in Section~\ref{sec:frontalization}, directly estimates a rigid motion and hence it preserves the facial expressions. The second method combines 3D-to-2D shape-to-image fitting with a style-transfer GAN model  \cite{zhou2020rotate}, denoted ST-GAN. The third one learns a non-linear image-to-image dual-attention GAN model \cite{yin2020dual}, denoted DA-GAN. We also consider the case of directly using the raw input without any form of face frontalization, denoted WithHM (with head movements).
In all these four cases we crop the lip region, which yields 67$\times$67 images, which are then converted to gray scale and normalized to facilitate the downstream processing.

We consider three speech enhancement models. The Audio-only VAE (A-VAE), \cite{Leglaive_MLSP18,sadeghi2020audio} has an encoder and a decoder composed of fully-connected (fc) layers. The extracted audio feature vector is of size $F=513$ whereas the latent space is of size $L = 32$. One audio-visual VAE model shares a similar encoder-decoder architecture as A-VAE, with the additional fully-connected layers to encode the visual information~\cite{sadeghi2020audio}. We denote this model as AV-CVAE. Furthermore, we propose to use the ResNet backbone from a pretrained model specialized for lip reading~\cite{martinez2020lipreading} (shown in a dashed box in Figure~\ref{fig:av-cvae}) for visual feature extraction. The backbone follows the standard design of ResNet-18~\cite{he2016deep} except for the first convolutional layer, which is replaced by a 3D convolutional layer to incorporate temporal information from neighbouring frames. This variant is denoted as Res-AV-CVAE. In practice, the dimension of the visual embedding is $M = 128$.

All the VAE models are trained in an end-to-end manner. The A-VAE model is trained on pure audio data. The pretrained model AV-CVAE \cite{sadeghi2020audio} is fine-tuned on the MEAD dataset, whereas Res-AV-CVAE models are trained from scratch. Note that the ResNet backbone is frozen without requiring the gradients. It is hence a static feature extractor. We set $5e^{-5}$ as the learning rate for the fine-tuning model and $1e^{-4}$ for the training from scratch. The optimizer is Adam and the batch size is of 128. We also applied early stopping with a patience of 10 epochs. Note that we trained and tested one model with one specific lip preprocessing method at a time. 
At test time, noise from the DEMAND~\cite{thiemann2013demand} dataset is combined with the clean speech to construct the audio input. For each noise type there are five noise levels: -10 dB, -5 dB, 0 dB, 5 dB and 10 dB. Three standard speech enhancement metrics are used for quantitative evaluation: scale-invariant signal-to-distortion ratio (SI-SDR)~\cite{le2019sdr}, short-time objective intelligibility (STOI)~\cite{taal2011algorithm} and perceptual evaluation of speech quality (PESQ)~\cite{rix2001perceptual}. SI-SDR is measured in decibels (dB), while STOI and PESQ values are in the range $\left[0,1\right]$ and $\left[-0.5,4.5\right]$, respectively (the higher the better).

\begin{figure}[t]
    \centering
    \includegraphics[width=\linewidth]{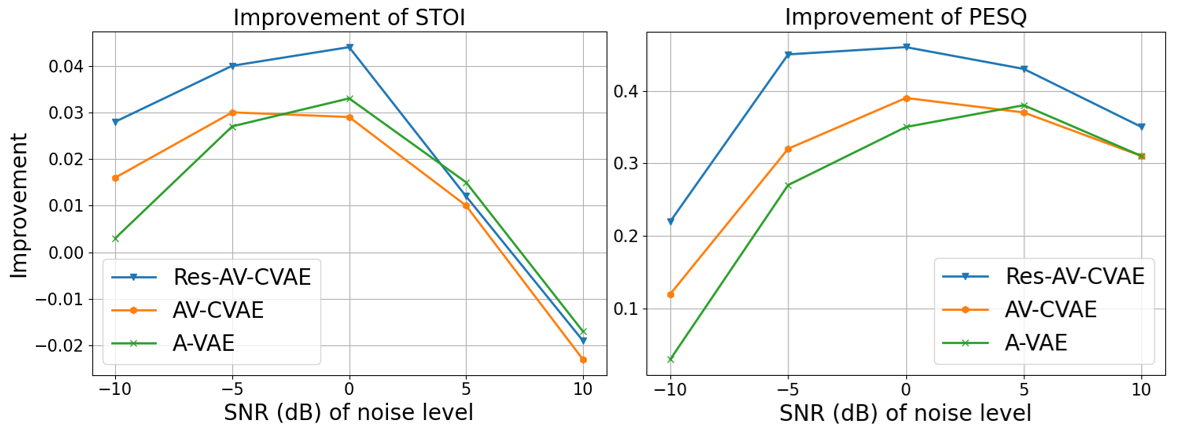}
    
    \caption{Performance comparison of A-VAE, AV-CVAE and Res-AV-CVAE based on STOI (left) and PESQ (right). }\label{fig:vae_compare}
    
 \end{figure}

We start with evaluating the impact of different frontalization methods on AVSE performance, i.e. Table~\ref{tab:measures}, where the average scores for different levels of noise (SNR) are presented. Selecting the RFF, the best-performing method in the table, as an example, we remark that the difference between with and without frontalization is significant. This confirms that the head motions interfere the capture of visual speech patterns. In other words, separating the rigid head movements from the non-rigid lip deformations allows the model to learn a better clean speech model. Moreover, the comparison between A-VAE and Res-AV-CVAE with RFF further validates the contribution of the visual modality. 

The choice of the face frontalization method is important. While RFF shows significant improvements, ST-GAN  yields a minor difference compared to the presence of head movements. Indeed, GAN-based image generation models have no theoretical guarantee for preserving the lip shape -- they add some form of visual noise, which neutralizes the gain of frontalization. This explanation is also supported by the results of DA-GAN: its performance is falling far behind the counterparts. As the results of ST-GAN are conditioned on the transformation-based process, the model possesses a prior knowledge about the frontalized face. Moreover, the direct mapping from an arbitrary viewpoint to a frontal view of DA-GAN introduces even more dramatic modifications in the lip shape. Thus, the model has more difficulties to learn the correct speech patterns from lip movements. 

We then compare the performance of different VAE architectures in Figure~\ref{fig:vae_compare}, where the improvement of scores are shown as a function of different levels of noise (SNR). More precisely, the improvement refers to the difference between the score obtained by using the raw noisy speech and that obtained by using the enhanced speech. First, it is remarkable to see that Res-AV-CVAE significantly outperforms AV-CVAE, showing the gain of using a more powerful feature extractor. Second, we observe that with a noise level at between -5 dB and 0 dB, the Res-AV-CVAE model reaches an optimal stage (a peak in the curve) for fusing the audio-visual data. That is, with the noise level going higher (smaller SNR), the audio would be too corrupted to be enhanced. In contrast, with the noise level going lower (higher SNR), the importance of the visual data is decreasing and the already clean speech becomes harder to be enhanced. While the superior performance of the Res-AV-CVAE models is more significant at high noise levels, it is quite remarkable to observe that Res-AV-CVAE-RFF performs almost equally well as A-VAE for low noise levels. These experiments confirm the 
complementary roles of the visual and audio modalities for the task of speech enhancement.

\begin{figure}[t]
\centering
\raisebox{0.6\height}{\includegraphics[trim = 15.5cm 3.5cm 14cm 12.5cm,clip,keepaspectratio=true,width=0.22\columnwidth]{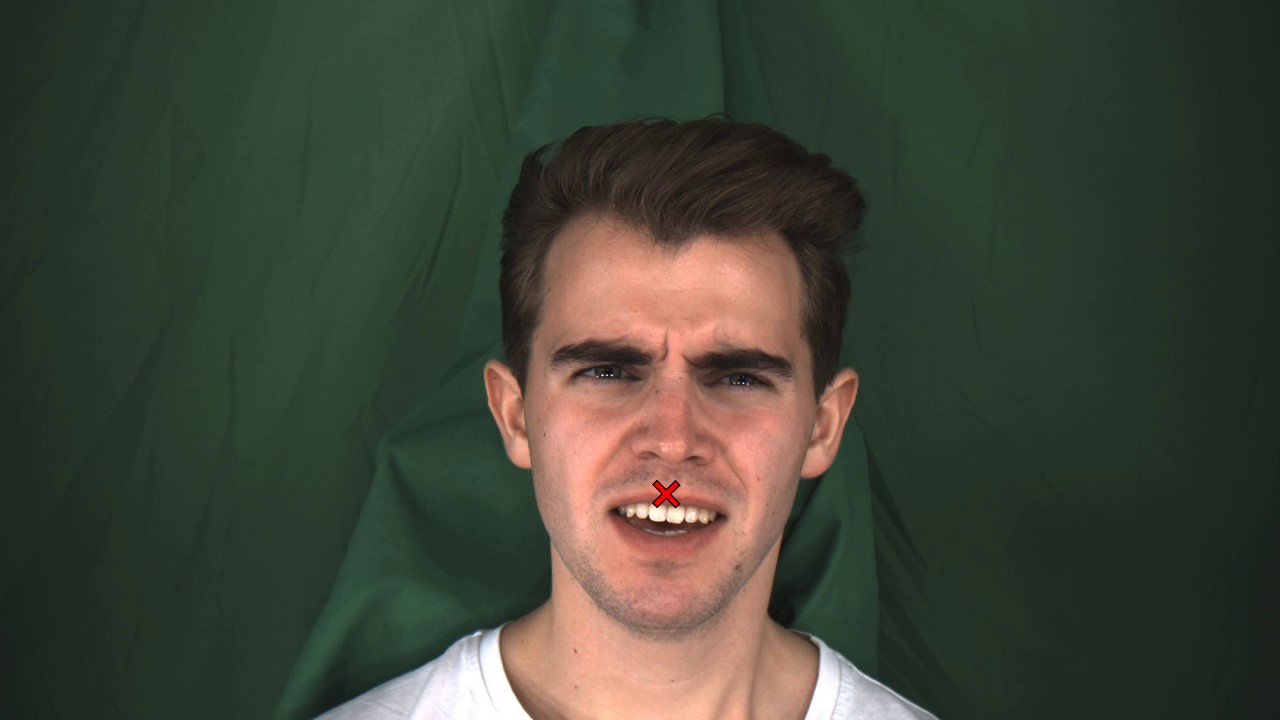}}
\includegraphics[trim = 2.5cm 0.4cm 4.5cm 0cm,clip,keepaspectratio=true,width=0.77\columnwidth]{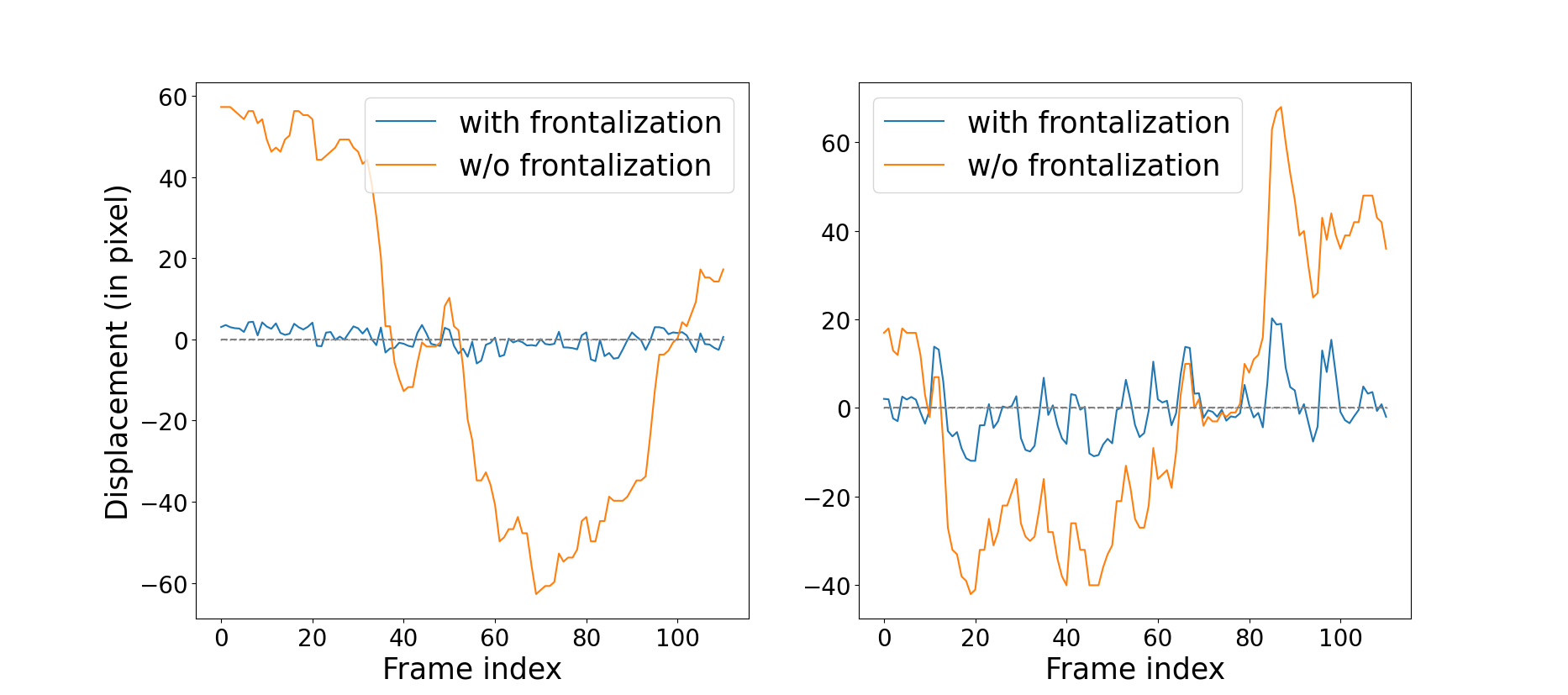}
\caption{Left: landmark location on the upper lip. Middle: horizontal landmark motion. Right: vertical landmark motion. Blue plots: Lip motions after frontalization (using RFF). Orange plots: lip motions in the presence of head movements  (without frontalization).}\label{fig:displacement}
\end{figure}

To give an insight on the impact of removing head movements, Figure~\ref{fig:displacement} shows the horizontal and vertical displacements of a landmark located on the upper lip. Both the vertical and horizontal trajectories of this lip landmark are strongly affected by head motions. In the light of this experiment, one may interpret the process of separating rigid head movements and non-rigid lip movements, as a way of extracting clean visual-speech information from the raw videos.

\section{Conclusion}
In this paper, we investigated the effect of head movements on the task of audio-visual speech. We showed that the combination of face frontalization for removing head movements in combination with a ResNet backbone considerably improves the performance of state-of-the-art VAE AVSE, based on several speech enhancement metrics. We compared a recently proposed RFF method \cite{kang2021robust} with two state-of-the-art frontalization methods, both based on GANs. We showed that the built-in expression-preserving property of \cite{kang2021robust} yields much better results than methods based on a GAN architecture -- the latter cannot guarantee that the frontalization process preserves the lip movements of the input.
In the future, we foresee supplementary experiments with large head movements and more extreme head poses. In addition, combining face frontalization with the recently proposed switching-VAE AVSE model \cite{sadeghi2021switching} is an interesting topic as well.

\setlength\bibitemsep{0.3em}
\printbibliography

@inproceedings{zhou2020rotate,
  title={Rotate-and-render: Unsupervised photorealistic face rotation from single-view images},
  author={Zhou, Hang and Liu, Jihao and Liu, Ziwei and Liu, Yu and Wang, Xiaogang},
  booktitle={CVPR},
  year={2020}
}

@inproceedings{yin2020dual,
  title={Dual-Attention {GAN} for Large-Pose Face Frontalization},
  author={Yin, Yu and Jiang, Songyao and Robinson, Joseph P and Fu, Yun},
  booktitle={IEEE International Conference on Automatic Face and Gesture Recognition},
  pages={24--31},
  year={2020},
}

@inproceedings{martinez2020lipreading,
  title={Lipreading using temporal convolutional networks},
  author={Martinez, Brais and Ma, Pingchuan and Petridis, Stavros and Pantic, Maja},
  booktitle={IEEE International Conference on Acoustics, Speech and Signal Processing},
  pages={6319--6323},
  year={2020},
 }

@inproceedings{le2019sdr,
  title={{SDR}--half-baked or well done?},
  author={Le Roux, Jonathan and Wisdom, Scott and Erdogan, Hakan and Hershey, John R},
  Booktitle = {ICASSP},
  year={2019},
 
}

@inproceedings{he2016deep,
  title={Deep residual learning for image recognition},
  author={He, Kaiming and Zhang, Xiangyu and Ren, Shaoqing and Sun, Jian},
  booktitle={Proceedings of the IEEE conference on computer vision and pattern recognition},
  pages={770--778},
  year={2016}
}

@inproceedings{thiemann2013demand,
  title={DEMAND: a collection of multi-channel recordings of acoustic noise in diverse environments},
  author={Thiemann, Joachim and Ito, Nobutaka and Vincent, Emmanuel},
  booktitle={Proc. Meetings Acoust},
  pages={1--6},
  year={2013}
}

@inproceedings{kang2021robust,
  title={Robust Face Frontalization For Visual Speech Recognition},
  author={Kang, Z. and Horaud, R. and Sadeghi, M.},
  booktitle={ICCV Workshops},
  year={2021}
}

@inproceedings{sadeghi2021switching,
  title={Switching variational auto-encoders for noise-agnostic audio-visual speech enhancement},
  author={Sadeghi, M. and Alameda-Pineda, X.},
  Booktitle = {ICASSP},
    year={2021},
 }

@inproceedings{wang2020mead,
  title={{MEAD}: A large-scale audio-visual dataset for emotional talking-face generation},
  author={Wang, Kaisiyuan and Wu, Qianyi and Song, Linsen and Yang, Zhuoqian and Wu, Wayne and Qian, Chen and He, Ran and Qiao, Yu and Loy, Chen Change},
  booktitle={ECCV},
  year={2020},
  organization={Springer}
}

@article{forbes2014new,
  title={A new family of multivariate heavy-tailed distributions with variable marginal amounts of tailweight: application to robust clustering},
  author={Forbes, F. and Wraith, D.},
  journal={Statistics and Computing},
  volume={24},
  number={6},
  pages={971--984},
  year={2014},
  publisher={Springer}
}

@techreport{kraft1988software,
  title={A software package for sequential quadratic programming},
  author={Kraft, D.},
  institution={DLR German Aerospace Center -- Institute for Flight Mechanics},
  address={Koln, Germany},
  year={1988},
  number={DFVLR-FB 88-28}
}

@inproceedings{bulat2017far,
  title={How far are we from solving the {2D} \& {3D} face alignment problem? (and a dataset of 230,000 3D facial landmarks)},
  author={Bulat, A. and Tzimiropoulos, G.},
  booktitle={ICCV},
  year={2017}
}

@article{sadeghi2020audio,
  title={Audio-Visual Speech Enhancement Using Conditional Variational Auto-Encoders},
  author={Sadeghi, Mostafa and Leglaive, Simon and Alameda-Pineda, Xavier and Girin, Laurent and Horaud, Radu},
  journal={IEEE/ACM Transactions on Audio, Speech, and Language Processing},
  volume={28},
  pages={1788--1800},
  year={2020},
  publisher={IEEE}
}

@InProceedings{paysan20093d,
  author       = {Paysan, Pascal and Knothe, Reinhard and Amberg, Brian and Romdhani, Sami and Vetter, Thomas},
  booktitle    = {IEEE International Conference on Advanced Video and Signal Based Surveillance},
  title        = {A {3D} face model for pose and illumination invariant face recognition},
  year         = {2009},
   pages        = {296--301},
}

@Article{taal2011algorithm,
  author    = {Taal, C. H. and Hendriks, R. C. and Heusdens, R. and Jensen, J.},
  title     = {An algorithm for intelligibility prediction of time--frequency weighted noisy speech},
  year      = {2011},
  volume    = {19},
  number    = {7},
  pages     = {2125--2136},
  journal   = {IEEE Trans. Audio, Speech, Language Process.},
  publisher = {IEEE},
}

@Conference{AfouCZ18,
  Title                    = {The Conversation: {D}eep Audio-Visual Speech Enhancement},
  Author                   = {Afouras, T. and Chung, J. S. and Zisserman, A.},
  Booktitle                = {INTERSPEECH},
  Year                     = {2018},
  Pages                    = {3244-3248}
}

@Conference{GabbSP18,
  Title                    = {Visual speech enhancement},
  Author                   = {Gabbay, A. and Shamir, A. and Peleg, S.},
  Booktitle                = {INTERSPEECH},
  Year                     = {2018},
  Pages                    = {1170--1174}
}

@Article{hou2018audio,
  Title                    = {Audio-visual speech enhancement using multimodal deep convolutional neural networks},
  Author                   = {Hou, J.-C. and Wang, S.-S. and Lai, Y.-H. and Tsao, Y. and Chang, H.-W. and Wang, H.-M.},
  Journal                  = {IEEE Transactions on Emerging Topics in Computational Intelligence},
  Year                     = {2018},
  Number                   = {2},
  Pages                    = {117--128},
  Volume                   = {2}
}

@Conference{KingW14,
  Title                    = {Auto-encoding variational Bayes},
  Author                   = {Kingma, D. P. and Welling, M.},
  Booktitle                = {ICLR},
  Year                     = {2014}
}

@InProceedings{Leglaive_MLSP18,
  Title                    = {A variance modeling framework based on variational autoencoders for speech enhancement},
  Author                   = {Leglaive, S. and Girin, L. and Horaud, R.},
  Booktitle                = {MLSP},
  Year                     = {2018}
}

@InProceedings{rix2001perceptual,
  Title                    = {{Perceptual evaluation of speech quality (PESQ)-a new method for speech quality assessment of telephone networks and codecs}},
  Author                   = {Rix, A. W. and Beerends, J. G. and Hollier, M. P. and Hekstra, A. P.},
  Booktitle                = {ICASSP},
  Year                     = {2001}
}

@Conference{SadeA19a,
  author    = {Sadeghi, Mostafa. and Alameda-Pineda, Xavier},
  title     = {Robust Unsupervised Audio-visual Speech Enhancement Using a Mixture of Variational Autoencoders},
  Booktitle = {ICASSP},
  year      = {2020},
}

@article{sadeghi2021mixture,
  title={Mixture of inference networks for {VAE}-based audio-visual speech enhancement},
  author={Sadeghi, Mostafa and Alameda-Pineda, Xavier},
  journal={IEEE Transactions on Signal Processing},
  volume={69},
  pages={1899--1909},
  year={2021},
  publisher={IEEE}
}

@Article{michelsanti2021overview,
  title={An overview of deep-learning-based audio-visual speech enhancement and separation},
  author={Michelsanti, D. and Tan, Z.-H. and Zhang, S.-X. and Xu, Y. and Yu, M. and Yu, D. and Jensen, J.},
  journal={IEEE/ACM Transactions on Audio, Speech, and Language Processing},
  year={2021},
  publisher={IEEE}
}

@Conference{afouras2019my,
  title={My lips are concealed: Audio-visual speech enhancement through obstructions},
  author={Afouras, Triantafyllos and Chung, Joon Son and Zisserman, Andrew},
  booktitle={INTERSPEECH},
  year={2019}
}

\end{document}